\documentclass{aastex63}
\usepackage{graphicx}
\usepackage{hyperref}
\usepackage{amsmath}

\submitjournal{\aj}

\shorttitle{ExoVista}
\shortauthors{Stark et al.}

\begin{document}

\title{ExoVista: A Suite of Planetary System Models for Exoplanet Studies}

\correspondingauthor{Christopher Stark}
\email{christopher.c.stark@nasa.gov}

\author{Christopher C. Stark}
\affiliation{NASA Goddard Space Flight Center \\
Exoplanets and Stellar Astrophysics Laboratory, Code 667 \\
Greenbelt, MD 20771, USA}

\begin{abstract}

Studies of future space- and ground-based exoplanet surveys often rely on models of planetary systems to simulate instrument response, estimate scientific yields, perform trade analyses, and study efficient observation strategies. Until now, no planetary system models contained all of the basic physics necessary to enable study with all of the major exoplanet detection methods. Here we introduce a suite of such models generated by a new tool, exoVista. The exoVista tool quickly generates thousands of models of quasi-self-consistent planetary systems around known nearby stars at scattered light wavelengths and efficiently records the position, velocity, spectrum, and physical parameters of all bodies as functions of time. The modeled planetary systems can be used to simulate surveys using the direct imaging, transit, astrometric, and radial velocity techniques, as well as the overlap of these different methods.

\end{abstract}


\section{Introduction} \label{sec:intro}

The design of astronomical observatories requires an understanding of the astrophysical phase space accessible to the instruments. In the field of exoplanet detection and characterization, this means an understanding of the possible planetary systems that could be studied. Exoplanet yield studies have been conducted for a broad range of exoplanet detection methods, including gravitational lensing \citep[e.g.,][]{penny2019}, the transit method \citep[e.g.,][]{barclay2018}, astrometry \citep[e.g.,][]{perryman2014}, nulling interferometry \citep[e.g.,][]{kammerer2018}, and direct imaging \citep[e.g.,][]{savransky2010}. These studies map models of the instrument response onto a set of discrete or statistical planetary system models to quantify the distribution of exoplanets that might be detected with a given survey, perform trade analyses that inform mission design, study efficient observation strategies, and optimize the potential target list. In the majority of cases, these analyses require modeling observations of planetary systems that have not yet been detected, hence the need for fictitious exoplanetary system models that reasonably represent what \emph{could} exist.

The planetary system models used for these studies are typically generated by combining exoplanet occurrence rates (sometimes extrapolated to unexplored phase space) with a stellar catalog. In addition, these studies must make a large number of assumptions about the optical (e.g., albedo and phase function), orbital (e.g., eccentricity distribution), and structural (e.g., mass-radius relationship) properties of the planets, as well as other aspects of the star (e.g., spectrum and variability) and planetary system (e.g., exozodi). Not surprisingly, these assumptions vary between studies, creating confusion when comparing studies. Additionally, these varying assumptions make it difficult to study the overlap of different exoplanet detection methods. As each of these exoplanet detection methods improves and explores a new region of phase space, a self-consistent set of planetary system models will become even more important.

While there are many existing tools that model exoplanets, stars, and debris disks, there are relatively few that attempt to produce a scene of the full planetary system for direct imaging simulations. The Haystacks models of \citet{roberge2017} focused on reproducing the appearance of the Solar System observed from afar. While these are the highest fidelity planetary system scene models yet, this tool is limited to a Solar System twin and requires hours to produce a single model that is $\sim$10 GB in size. The ExoScene code was used to produce exoplanetary scene simulations for the successful Roman Coronagraphic Instrument (CGI) Data Challenge \citep{turnbull2021}, but relies on a limited set of external debris disk models. Similarly, the SISTER code of \citet{hildebrandt2021} incorporates starshade optical modeling with an exoplanet scene generation tool and was used for the Starshade Exoplanet Data Challenge, but relies on a user-supplied debris disk model. Importantly, none of these exoplanet scene generation tools produce self-consistent planet and dust architectures for general planetary systems, none track the gravitational interactions between planets, and none attempt to reproduce the statistical distributions of exoplanets or interplanetary dust from measured occurrence rates.

Here we describe a new tool, exoVista, that produces a ``universe" of planetary systems. ExoVista quickly produces a large number of models that can serve as a standard set of planetary systems for simulated study with the direct imaging, transit, astrometric, interferometric, and radial velocity (RV) methods at scattered light wavelengths. ExoVista distributes planets consistent with the \emph{Kepler} occurrence rates around \emph{Hipparcos} stars within 50 pc \citep{dulz2020}, assigns a mass, radius, and albedo to each planet, checks for stability of the orbits, evolves all objects with a gravitational n-body integrator, and generates a quasi-self-consistent debris disk for each system consistent with the LBTI HOSTS exozodi survey \citep{ertel2020}. The end result is a model that contains the basic physics needed for cursory study with the majority of exoplanet detection techniques. The outputs of exoVista can be used to simulate full exoplanet surveys of a population of stars, allowing us to study, e.g., the overlap of different exoplanet detection methods, the benefits of prior knowledge, and how to optimize the decision-making process during a survey. Additionally, we can use the scenes from exoVista to simulate direct imaging observations in detail, including the spectral extraction of exoplanets, the impact of exozodiacal dust on exoplanet detection and characterization, and the identification of planets between multiple epochs of observation.

For this initial release of exoVista, written in IDL and C, we have pre-computed a dozen ``universes" of exoplanetary systems around stars within 30 pc, all available for download via the Goddard EMAC web server at \url{https://ckan-files.emac.gsfc.nasa.gov/minio/exovista/}. We note that exoVista is a platform on which we intend to build. Future versions will incorporate more complete stellar catalogs, additional planet spectra, stellar variability, and phase-dependent planet spectra, among other things. Below we describe the basic functionality of the initial release. Section \ref{sec:methods} describes the main modules of the exoVista software and the methods and assumptions we adopt for each module. Section \ref{sec:results} presents example results and describes how one can probe the data set using different exoplanet detection methods.

\section{Methods \& Modules} \label{sec:methods}

Three main goals define the philosophy behind the design of exoVista. First, we designed exoVista to rapidly generate astrophysical scenes so that it can be applied to a large number of stars. By achieving numerical run times $\sim$minutes per target per core, we can generate thousands of planetary scenes in one day on a reasonable server.

Second, we designed exoVista to minimize the file size of the models. By achieving file sizes $\sim$10 MB per star, we can store thousands of astrophysical scenes in $<$1 TB of disk space, and substantially reduce load times. To achieve these reduced file sizes we limit the tool to scattered light wavelengths and save the debris disk data cube contrast, as opposed to surface brightness. The former has smoothly varying, broad spectral features, allowing for coarse spectral sampling, while the latter has the imprint of the stellar spectrum on it and requires finer spectral sampling.

Finally, the outputs of exoVista are broad enough to be useful to a wide range of planet detection techniques. Self-consistent barycentric and heliocentric positions and velocities of the star and planets are recorded as functions of time, as well as each object's spectrum, allowing for investigation with the RV, astrometric, and direct imaging methods. By comparing the heliocentric coordinates with stellar radius, the data set can be applied to the transit method. Further, the transit time can be approximated for transit timing variation (TTV) analyses.

The exoVista code is split into four sequentially-called modules: {\tt load\_stars}, {\tt generate\_planets}, {\tt generate\_disks}, and {\tt generate\_scene}. Here we describe the functionality of each module and the assumptions made within. Table \ref{table:parameters} provides a summary of the parameters used within exoVista.

\subsection{{\tt load\_stars}}

The first module of exoVista, {\tt load\_stars}, loads the target list. Currently {\tt load\_stars} reads the $\sim$8000 stars within 50 pc from the HabEx and LUVOIR master target list \citep{stark2019}. The HabEx/LUVOIR master target list was generated by combining the \emph{Hipparcos} \citep{perryman1997,vanleeuwen2007} and \emph{Gaia} TGAS/DR2 \citep{gaiadr1,marrese2018} target lists within 50 pc. All \emph{Hipparcos} distances were updated with \emph{Gaia} distances. The Washington Double Star (WDS) catalog was referenced to obtain simple binary parameters (separation and differential magnitude of the nearest companion); no other binary catalogs were referenced. Spectral types and photometry were obtained from the original \emph{Hipparcos} release and updated with SIMBAD.

We remove all luminosity class III and IV stars, leaving only main sequence (MS) and sub-giant stars, along with unclassified luminosity classes that we assume are predominantly MS. We then estimate stellar angular diameters using the B-V analytic fits from \citet{boyajian2014}; we calculate physical stellar diameters from the angular diameter and distance. We use the MS stellar properties table from \citet{pecaut2013} to estimate stellar mass and effective temperature by interpolating B-V. We then use the estimated mass and stellar radius to estimate $\log{g}$. Finally, we estimate luminosity from the absolute V band magnitude and a color correction using Equation 9 of \citet{torres2010} and the color correction factors in footnote 10 of \citet{pecaut2013}. 

We emphasize that this target list is not intended to be accurate on a star-by-star basis. In addition to containing occasionally inaccurate stellar parameters, many M stars are missing due to the $\sim$8$^{\rm th}$ magnitude limit of the \emph{Hipparcos} catalog, and we expect a non-negligible fraction of these stars to have unknown binary companions. The current input catalog is intended to roughly represent the \emph{population} of nearby MS and sub-giant stars brighter than 8$^{\rm th}$ magnitude. This target list could be substantially improved in accuracy and completeness, but a thorough analysis of every star in this list is beyond the scope of this project. While extending the target list to include more M stars should have a negligible impact on future direct imaging mission simulations that target potentially Earth-like planets, it would significantly expand the applicability of this tool to transit survey simulations.

We note that exoVista is not limited to the target list generated by {\tt load\_stars}. Users may supply their own target list, as long as format of the data structure is consistent and a minimum set of parameters are provided.

{\tt load\_stars} returns an $n$-element array of structures, $s$, where $n$ is the number of stars. Each entry of $s$ contains a stellar ID number, \emph{Hipparcos} ID, Tycho ID, right ascension, declination, UBVRIJHK photometric magnitudes, absolute V magnitude, distance, spectral type, bolometric luminosity, effective temperature, angular diameter, physical diameter, mass, $\log{g}$, and WDS separation and differential magnitude of the nearest companion star. Stellar spectra are determined later, in the {\tt generate\_scene} module.

\subsection{{\tt generate\_planets}}

After the target list is loaded, we use the {\tt generate\_planets} module to distribute planets among stars. {\tt generate\_planets} modifies the structure array $s$ to include a system midplane inclination, $i$, and position angle (PA). We draw inclinations randomly from a distribution uniform in $\cos{i}$ and position angles uniformly from $[0,\pi)$.

We then randomly assign planets to stars. To do so, we use the occurrence rate map in mass-semi-major axis space for FGK stars from \citet{dulz2020}. \citet{dulz2020} extrapolate the \emph{Kepler} and RV occurrence rates to smaller masses and longer periods, while using dynamical stability arguments to limit the number of planets at large distances. Importantly, we apply the occurrence rate map in $(M,a_{L_{\star}})$ space, where $M$ is planet mass and $a_{L_{\star}}$ is the semi-major axis multiplied by the square root of the stellar bolometric luminosity. By doing so, the occurrence rate maps scale with the habitable zone such that $\eta_{\rm Earth}$ is independent of spectral type. We treat the occurrence rate map as a probability distribution. For all $n$ stars we perform a Monte Carlo draw on every bin of the occurrence rate map to find the total number of planets to distribute to all stars in each bin. As a result, when running exoVista the number of planets generated may vary from run to run and be greater or less than the expectation value. We note that the occurrence rates of \citet{dulz2020} have significant uncertainties at large periods. These uncertainties must be kept in mind when applying the astrometric or direct imaging techniques to the exoVista dataset. To help estimate the impact of this uncertainty, we provide additional occurrence rate maps that correspond to the 1-$\sigma$ upper and lower limits.

After determining the total number of planets to generate for each $(M,a_{L_{\star}})$ bin, we randomly assign those planets to stars. We assume that planet mass and semi-major axis is randomly spaced logarithmically within each $(M,a_{L_{\star}})$ bin. We use planet mass to determine radius via the relationship of \citet{chenkipping2017}, given by
\begin{equation}
	R = 
	 \begin{cases}
		1.008\left(M/1.00\right)^{0.279} & \text{for $M < 2.04$ $M_{\rm Earth}$}\\
		1.230\left(M/2.04\right)^{0.589} & \text{for $2.04 \le M \le 131.6$ $M_{\rm Earth}$}\\
		14.31\left(M/131.6\right)^{-0.044} & \text{for $M > 131.6$ $M_{\rm Earth}$}
	\end{cases}
\end{equation}
where $R$ is the radius of the planet. Currently we assume a singular mass-to-radius conversion with no spread in the distribution. A realistic distribution of radii for each mass, which we leave for future work, would lead to broader range of possible planet fluxes and transit depths. As long as that distribution is symmetric, we do not expect it to significantly impact future mission studies in the limit of large target sample sizes.

By default, we set all orbital eccentricities to zero, as was also assumed by \citet{dulz2020} when estimating the impact of dynamical stability constraints on occurrence rates. However, the user can specify a non-zero range and the orbital eccentricities will propagate through the rest of the code self-consistently (e.g., when determining stability, the debris disk distribution, and integrating orbits).  Eccentric orbits would lead to greater spacing between simulated planets as well as larger peak stellar radial velocity signals near periapse passage. We also assume planetary orbital inclinations are uniformly distributed from $0$--$5^{\circ}$ by default, relative to the system midplane. All other orbital angles are randomly distributed between $[0,2\pi)$. Our orbital assumptions effectively exclude the presence of eccentric, highly-inclined hot Jupiters. While such exoplanets are rare, studies using the radial velocity and transit methods, which are biased toward finding such planets, should not expect the dynamical aspects of this population to be represented within the exoVista dataset.

After adding planets, we check all planet pairs to determine if they are dynamically stable. To do so, we take the same approach as \citet{dulz2020} and calculate the mutual Hill radius, given by
\begin{equation}
	\Delta = 2 \left(\frac{a_{\rm out} - a_{\rm in}}{a_{\rm out} + a_{\rm in}}\right) \left(\frac{3 M_{\star}}{M_{\rm out} + M_{\rm in}}\right)^{1/3},
\end{equation}
where $a_{\rm in}$ and $M_{\rm in}$ are the semi-major axis and mass of the inner planet, $a_{\rm out}$ and $M_{\rm out}$ are the semi-major axis and mass of the outer planet, and $M_{\star}$ is the mass of the host star. We require $\Delta > 6$ and delete all planets that don't meet this criterion. We then iteratively add more planets until the desired number per $(M,a_{L_{\star}})$ bin are generated. If any bin has not achieved its desired number of planets in 50 iterations, the creation of planets ends. We note that the occurrence rates of \citet{dulz2020} adopted $\Delta > 9$ to ensure Gyr of stability. Adopting the same criteria here greatly slows the planet distribution process. Because we are generating planetary systems that are \emph{plausible}, and not necessarily strictly stable for many Gyr, we relaxed this criterion to substantially speed up the distribution of planets. We note that we do not currently attempt to reproduce the empirically measured planet multiplicity, nor do we include mean motion resonant configurations of exoplanets.

We then randomly assign each planet a wavelength-dependent geometric albedo based on simple rules applied to each $(M,a_{L_{\star}})$ bin. For the geometric albedos, we currently adopt the geometric albedos included in the Haystacks project \citep{roberge2017}, which includes models of Venus, Earth, Mars, Jupiter, Saturn, Uranus, and Neptune. Briefly, the terrestrial planet albedos were derived from reflectance calculations by the Virtual Planet Laboratory using the Spectral Mapping Atmospheric Radiative Transfer (SMART) model \citep{meadows1996}. The disk-integrated reflectance of Earth was derived assuming an equatorial view averaging over the daily rotation, but can be applied to a broad range of viewing angles with $\sim30\%$ accuracy. The albedos of the giants planets were derived from the $300$-$900$ nm albedo spectra of \citet{karkoschka1998}. The $900$-$2500$ nm SpeX observations of \citet{rayner2009} were used to cover the near infrared after dividing by the solar spectrum and scaling to match the albedo spectra of \citet{karkoschka1998}. Further details of how these albedo were determined, as well as plots of their wavelength dependence, can be found in \citep{roberge2017} and references therein. We emphasize that the exoVista model albedo are not intended to be used to fit to observations. Rather, they are intended to loosely represent the variety of exoplanet albedo that may occur in nature and serve as a known input for exoplanet spectral retrieval studies. We assign each albedo file an integer probability from 1--100 that determines the relative likelihood of drawing that planet in the given $(M,a_{L_{\star}})$ bin.

We treat the phase space of exoEarth candidates (EECs) separately to more easily control the distribution of planets within this region. We adopt the same definition of EECs as in the HabEx and LUVOIR final reports \citep{habex_final_report,luvoir_final_report}. All EECs are assigned some form of an Earth-like atmosphere (Archean, Hazy Archean, Proterozoic High-O$_2$, Proterozoic Low-O$_2$, and Modern Earth) based on the approximate relative timescales of these phases during Earth’s history \citep{arney2016}.

The result of {\tt generate\_planets} is a modification to the structure array $s$. In addition to containing the midplane inclination and PA for each system, {\tt generate\_planets} adds an additional structure array $p$ to each entry of $s$. $p$ is a 30-element array containing each existing planet's parameters: mass, radius, semi-major axis, eccentricity, inclination, longitude of ascending node, argument of periastron, initial mean anomaly, and assigned albedo file. For a target list of $\sim$8000 stars, {\tt generate\_planets} typically takes a total of $\sim$10 s to run.

\subsection{{\tt generate\_disks}}

After the planets are distributed around each star, we use the module {\tt generate\_disks} to quasi-self-consistently generate parameters defining the debris disk of each star based on the underlying planetary system. Here we describe the debris disk generation process in detail, as this is a novel model developed specifically for exoVista.

Observations of debris disks have revealed several common aspects about their geometry that our models attempt to replicate. For a summary of observed debris disk geometries, see \citet{hughes2018}. First, observed disks, which are hundreds to thousands of times denser than the Solar System zodiacal cloud, commonly appear as circumstellar rings. These rings have been shown to be relatively well fit with radial Gaussians at mm wavelengths, suggesting a well-constrained population of larger grains that track their planetesimal parent bodies. At short wavelengths we observe smaller grains, which are significantly impacted by radiative forces. The small grains create an extended halo of dust beyond the circumstellar ring to large distances, as they are blown onto eccentric orbits with large semi-major axes by radiation pressure. Radiative drag forces also cause these small dust grains to spiral inward from the circumstellar ring. For the massive, collisional disks we currently observe, only a small fraction of the dust spirals inward, such that the disk density interior to the ring is a much smaller fraction of the ring itself. However, theoretical studies suggest a much larger fraction of dust will spiral inward in less massive disks where collisions are rare, so much so that the disk can appear to be continuous all the way from the ring in to the sublimation distance when viewed at visible wavelengths \citep[e.g.,][]{wyatt99,kuchnerstark2010,lohne2012,kennedy2015}.

Observations have also revealed that debris disks commonly have a cold component at tens of AU, and a warm component at a few AU \citep{hughes2018}. By default, we distribute two disk components (warm and cool) to each system, though a maximum of three are possible if desired by the user. To distribute the dust, we first determine plausible locations for the circumstellar ring of parent bodies based on the underlying planetary system. To do so, we determine all semi-major axes that are not within 3 Hill radii of a planet. For the inner warm component, we find the most stable region (defined as the largest width of stable semi-major axes in log space) between 0.5 and 5 AU. For the outer cool component, we repeat the process from 5--50 AU. If the user desires a third cold disk component, we repeat the process again from 50--500 AU.

\begin{figure}[h]
	\centering	
	\includegraphics[width=4in]{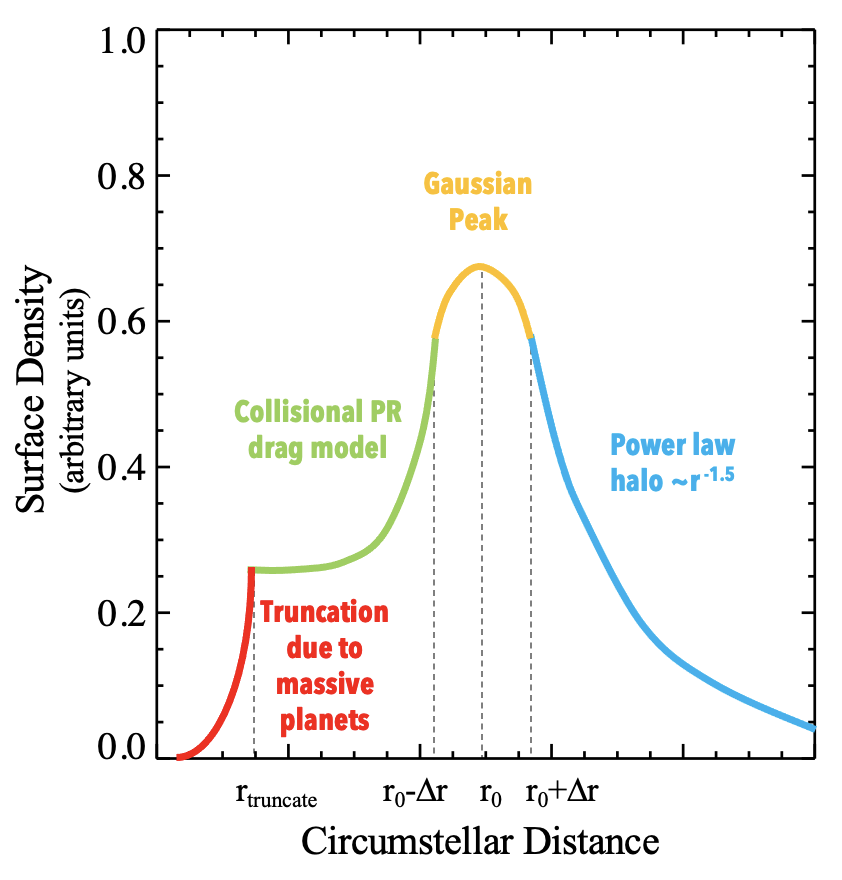}
	\caption{Illustration of the piece-wise analytic dust distribution model used for each disk component in {\tt generate\_disks}.\label{fig:diskmodel}}
\end{figure}

We model each disk component using a piece-wise combination of analytic models as shown in Figure \ref{fig:diskmodel}. This includes a Gaussian parent body ring of half-width $\Delta r$ centered at circumstellar distance $r_0$ (yellow line in Figure \ref{fig:diskmodel}). The width of the Gaussian ring ranges from $[0.05,0.3]r_0$, and is assigned based on the smaller of the range maximum (0.3) and the maximum width of the stable region at $r_0$. Only stable regions wider than the range minimum (0.05) are considered. Within the Gaussian ring we adopt a Dohnanyi size distribution \citep{dohnanyi1969}, given by
\begin{equation}
	\frac{dN}{ds} \propto s^{-3.5}. \label{eq:dohnanyi}
\end{equation}

Outside of $r_0+\Delta r$ (blue line in Figure \ref{fig:diskmodel}) we model the disk as an $r^{-1.5}$ power law dust halo. Although the outer halo portion should have a non-Dohnanyi size distribution \citep{thebault2014}, the expected size distribution is not well-described analytically, nor well-constrained observationally; we currently assume a Dohnanyi size distribution for the halo as well. The size distribution of our optically thin disk models mostly impacts the color of the disk, as the surface brightness was normalized to that expected for one ``zodi." A steeper-than-Dohnanyi size distribution, with an enhanced population of small grains, would appear bluer, whereas a flatter size distribution would appear redder, which could impact exoplanet spectral retrievals.

Interior to $r_0-\Delta r$ (green line in Figure \ref{fig:diskmodel}) we model the the disk as inward-migrating, colliding dust. We use Equation 3.36 from \citet{wyatt99}, which accounts for the collisional destruction of dust grains via
\begin{equation}
	\Sigma(r) = \frac{\Sigma(r_0)}{1 + 4 \eta_0 (1 - \sqrt{r/\left(r_0-\Delta r\right)})},
\end{equation}
where $\Sigma$ is the surface density, $\eta_0 = t_{\rm PR}(r=r_0-\Delta r)/t_{\rm coll}(r=r_0-\Delta r)$, $t_{\rm PR}(r)$ is the Poynting-Robertson (PR) drag time evaluated at circumstellar distance $r$ \citep{wyattwhipple1950}, and $t_{\rm coll}$(r) is the collisional timescale evaluated at $r$. We approximate the collision time for each disk as
\begin{equation}
	t_{\rm coll} = \frac{t_{\rm orbit}}{4 \pi \left(z \times 10^{-7}\right)},
\end{equation}
where $z$ is the number of zodis assigned to the disk and $t_{\rm orbit}$ is the orbital period. We calculate the PR drag time \citep{wyattwhipple1950} individually for each grain size in the disk, such that the inward-migrating portion can have a significantly non-Dohnanyi size distribution for massive disks. As a result, the inner regions of the disk can exhibit a radial color gradient.

Systems that host massive planets can dynamically eject dust \citep[e.g.,][]{moromartin2002}. To include this effect, we multiply the dust distribution by a factor of $(r/r_{\rm truncate})^3$, where $r_{\rm truncate}$ is set to $1.1$ times the apastron distance of the outer-most planet interior to the belt that has a mass $>$100 Earth masses (red line in Figure \ref{fig:diskmodel}). This truncation factor is motivated from the modeled optical depth reduction of inward-migrating Kuiper Belt dust due to Jupiter and Saturn \citep{kuchnerstark2010}.

The density of each analytic piece is scaled to ensure continuity at the transitions. All analytic pieces of a given disk component are randomly assigned the same scale height $H/r_0$ ranging from [0.03,0.2]. The vertical variation of each disk component's density is modeled by
\begin{equation} \label{eq:nvz}
 n\left(r,z\right) = \frac{n\left(r\right)}{r} e^{-0.5 \left({z}/{\left(r \left(H/r_0\right) \right)}\right)^2},
\end{equation}
where $n(r)$ is the radial density distribution, $z$ is the distance normal to the disk component's midplane, and the $1/r$ factor results from normalization of the vertical distribution. 

Finally, each disk component is assigned a scattering phase function (SPF). Observations have revealed that the shape of debris disks' SPFs can be successfully described using a linear combination of Henyey-Greenstein (HG) functions \citep{stark2014,sai2015,milli2017,matthews2017,chen2020}, expressed as
\begin{equation}
	p\!\left(g,\theta\right) = \sum_i \frac{w_i}{4\pi}\frac{1-g_i^2}{[\,1+g_i^2-2g_i\cos{\theta}\,]^{3/2}},
\end{equation}
where $\theta$ is the scattering phase angle, $w_i$ and $g_i$ are the weight and asymmetry parameter of the $i^{\rm th}$ HG function, respectively, $\sum_i w_i = 1$, and $g_i$ ranges from -1 for perfect backscattering to 1 for perfect forward scattering \citep{henyey1941}.  Unfortunately, because we cannot access the full range of scattering angles of any observed debris disk, we cannot know for sure that fits to observations remain valid at all scattering angles \citep{hedmanstark2015}. Critically, this lack of knowledge creates an uncertainty in the normalization of measured SPFs, making their application to models challenging.

\begin{figure}[h]
	\centering	
	\includegraphics[width=4in]{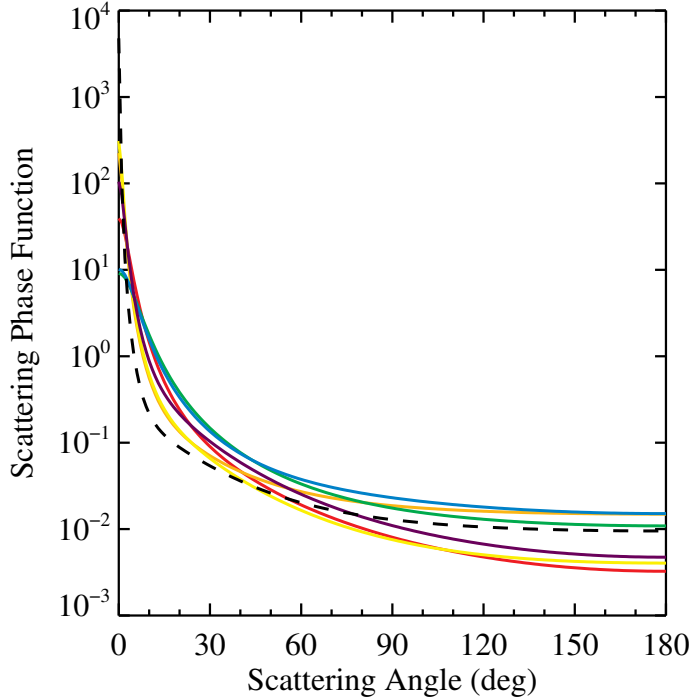}
	\caption{A sample of the variety of scattering phase functions produced by {\tt generate\_disks} (colored lines) compared to the best fit to Saturn's D68 ringlet of \citet{hedmanstark2015} (black dashed line).\label{fig:spfvariety}}
\end{figure}

In lieu of a measured, properly-normalized SPF for debris disks, we chose to use the next best thing: the SPFs of Saturn's optically thin G and D ringlets, which serve as a good proxy for debris disks, have been measured over nearly the full range of scattering angles, and can be properly normalized with reasonable accuracy \citep{hedmanstark2015}. \citet{hedmanstark2015} showed that a linear combination of three HG functions with $w_i = [0.754, 0.151, 0.095]$ and $g_i = [0.995, 0.585, 0.005]$ fit the measured SPFs well. However, simply adopting this single best fit would not properly sample the variety of SPFs that has been observed in debris disks \citep{hughes2018}. Therefore, we opt to randomly vary the weights and asymmetry parameter of each of the three HG functions. Specifically, we randomly and uniformly select values of $0.85 \le g_1 \le 0.98$, $0.5 \le g_2 \le 0.7$, and $-0.2 \le g_3 \le 0.2$ for the asymmetry parameters, and $0.6 \le w_1 \le 0.9$, $0.1 \le w_2 \le (1.0-w_1)$, and $w_3 = 1.0-w_1-w_2$ for the weights. This variation produces the SPFs shown in Figure \ref{fig:spfvariety}, which sample a variety of slopes and values near $\theta=90^\circ$, where most debris disk dust is observed. This variety will alter the degree of forward scattering in individual disks, as well as the apparent albedo, while maintaining an average apparent albedo consistent with \citet{hedmanstark2015}. While we allow the SPF to vary between disk components, we do not currently vary it with dust grain size.

The density of the inner warm disk is drawn randomly from the free form distribution that best fits the LBTI HOSTS survey results, which has a median of 3 zodis of dust \citep{ertel2020}. The outer cool disk is randomly assigned a density within a factor of 5 of the inner disk, in a basic effort to roughly correlate coeval disks. The uncertainties on the LBTI HOSTS best fit distribution are significant and can impact studies of future direct imaging missions, as brighter disks make it more difficult to detect the faint planets. To allow the user to constrain this impact, we include alternative distributions that represent the 1-$\sigma$ upper and lower limits from the LBTI HOSTS results \citep{ertel2020}. Importantly, we introduce an additional normalization factor to account for the SPF. We normalize the density of the disks such that a solar system twin model with our average SPF, inclined by 60 degrees from face-on, produces a V band surface brightness of 22 mag arcsec$^{-2}$ at a separation of 1 AU and scattering angle of $90^\circ$. This defines 1 ``zodi" of dust using the same convention as the HabEx and LUVOIR Final Reports \citep{habex_final_report,luvoir_final_report}.

We note that our disk models currently ignore debris disk asymmetries, including clumpy mean motion resonant structures \citep[e.g.,][]{jacksonzook1989} as well as warps and offsets due to secular perturbations from massive planets \citep[e.g.,][]{dermott1995}. These asymmetries could have significant effects on future directing imaging methods, by mimicking or obfuscating planets \citep[e.g.,][]{defrere2012}. While future versions of exoVista may include some of these effects, we emphasize that the impact of these effects on observations must currently be examined separately.

{\tt generate\_disks} modifies the structure array $s$ to include a new three element array structure $disk$ for each element of $s$. Each entry of $disk$ includes all of the disk's parameters described above. For a target list of $\sim$8000 stars, {\tt generate\_disks} typically takes a total of a few seconds to run.

\subsection{{\tt generate\_scene}}

Finally, after all planet and disk parameters are generated for each star, we use the module {\tt generate\_scene} to model the astrophysical scenes. We first define a wavelength array for the star and planets ($0.3$--$1.0$ $\mu$m with spectral resolution $R=300$ by default). We also define the wavelength array for the disk, which defaults to the same range as the star and planets, but with reduced spectral resolution $R_{\rm disk}=10$. We can use a lower spectral resolution for the disk because the disk’s spectral features are broad at visible wavelengths. The lower spectral resolution allows us to save substantial disk space for the output products. While the user can modify the wavelength array for any object, we emphasize that exoVista does not currently handle thermal emission. 

Next, we define the range of dust grain sizes to include in the debris disk model. We set the maximum grain size to $100\times$ the longest wavelength desired, which ensures that contributions from large grains do not contribute appreciably to the image. We set the minimum grain size to the blowout size, below which dust grains are removed from the system on short timescales by radiation pressure. While the blowout size is believed to vary with spectral type, we currently set the blowout size to 0.5 $\mu$m for all stars, as the blowout size of individual disks is poorly understood and depends strongly on the dust composition and porosity \citep{lebreton2012}. We also set the dust grain size resolution, $s/\Delta s$. By default $s/\Delta s = 5$, which we have tested to produce adequate results using a wide range of grain compositions and size distributions. 

We then loop over each star and generate its model outputs. For each star, we first retrieve an approximate stellar spectrum \citep{castelli2004} based on the stellar effective temperature, luminosity, and log(g). We normalize the stellar spectral model to reproduce the star’s observed V band flux. We do not normalize the stellar spectrum to the bolometric luminosity or radius, which are approximate quantities derived from analytic fits to B-V for main sequence stars. We also do not attempt to fit the stellar spectrum to any additional observed photometry, and thus reddening due to extinction is not included.

Next, we distribute dust grains based on the geometry and parameters of each disk component and generate a contrast image of each disk component. We do this on a pixel-by-pixel basis over the full image of the disk, solving the flux in each pixel to a predetermined tolerance. By default, the pixel scale is 2 mas and the field of view is 0.5$\arcsec$. The tolerance per pixel $\eta$ is set to $0.05/z$ and limited to a range of $[0.0005,0.05]$. This tolerance is roughly equivalent to 1/3 of a $\Delta {\rm mag}=26.5$ point source observed with a 4 m telescope, the assumed $3\sigma$ noise floor for HabEx. We limit $\eta$ to a minimum of $0.0005$ under the assumption that disks brighter than one hundred zodis will not be probed via direct imaging for planets with $\Delta {\rm mag}=26.5$, consistent with the yield simulations for the HabEx mission \citep{habex_final_report}. Because pixels near the star dominate the run time, we set $\eta = 0.1$ for all pixels within 15 mas of the star, equivalent to $3.5\lambda/D$ for a 15 m telescope at a wavelength of 300 nm; we do not expect any future direct imaging mission to routinely access circumstellar distances this small and are justified in relaxing the precision of disk flux calculations at these locations.

To calculate the disk flux in each pixel in the image, we generate an array of coordinates sub-sampling $(x',y',z')$ space within that pixel, where $(x',y')$ sample the East-West and North-South directions, respectively, $z'$ samples along the line of sight in the image, and $z'=0$ at the distance to the target star. Samples in the $(x',y')$ dimensions are uniformly spaced, whereas samples in the $z'$ dimension are spaced $\propto z'^2$ to achieve better resolution at the disk midplane and small circumstellar distances. We set the maximum value of $z'$ along our line of site to $r_{0,{\rm max}} \eta^{-2/7}$, where $r_{0,{\rm max}}$ is the maximum value of $r_0$ of all disk components. This ensures that $z'$ extends to a distance at which the surface brightness falls to a fraction $\eta$ of the brightness at $r_{0,{\rm max}}$, based on the $r^{-1.5}$ dust halo density distribution and an $r^{-2}$ illumination factor.

For each pixel, we start by sub-sampling the pixel using $n_{\rm xy}=1$ points in the $(x',y')$ dimensions and $n_{\rm z}=128$ points in the $z'$ dimension. We first resolve the disk in the $z'$ dimension by doubling $n_{\rm z}$ until the total flux in the pixel varies by less than the set tolerance. We then resolve the disk in the $(x',y')$ dimensions by doubling $n_{\rm xy}$ until the total flux varies by less than the set tolerance. The final number of points used in a given pixel is equal to $n_{\rm z,f} \times n_{\rm xy,f}^2$, where $n_{\rm z,f}$ and $n_{\rm xy,f}$ are the final resolutions in the $z'$ and $(x',y')$ dimensions, respectively.

At each sub-sampled point, we calculate the disk density of all three disk components based on the parameters generated in the module {\tt generate\_disks}. We first transform the sub-sampled $(x',y',z')$ points to the disk coordinates $(x,y,z)$, then evaluate the piece-wise analytic radial functions illustrated in Figure \ref{fig:diskmodel}. If a massive planet is present, we include the truncation model shown in red. We then multiply by the vertical variation given by Eq. \ref{eq:nvz}. 

Next, we multiply this density by the dust grain size distribution, scattering cross section, and scattering efficiency, and then integrate over all grain sizes. Dust grain scattering efficiencies are determined via Mie theory calculations. Performing Mie theory calculations on the fly would greatly slow the calculations, so we pre-calculated scattering efficiencies for specific grain sizes at the default size resolution $s/\Delta s=5$. For each discrete value of $s$ representing a bin of sizes of width $\Delta s$, we calculated the scattering efficiencies by averaging over 100 sub-sampled grain sizes assuming the Dohnanyi size distribution given by Equation \ref{eq:dohnanyi} within the sub-sample. This sub-sampling is necessary to reduce ``ringing" artifacts from Mie theory. We note that the assumption of a Dohnanyi distribution over our sub-sampled $\Delta s$ does not limit us to a Dohnanyi distribution overall; tests show reasonable agreement between our sub-sampled method and a high resolution calculation even for size distributions as steep as $dN/ds\propto s^{-5.5}$. Currently scattering efficiencies have only been calculated for astronomical silicates \citep{li2001}; future versions may include additional dust compositions.

Finally, we multiply by the SPF of the dust, as given by the linear combination of three Henyey-Greenstein functions generated by {\tt generate\_disks}, the differential volume of each sub-sampled $(x',y',z')$ point, and sum over all points to determine the disk contrast per pixel (flux of disk per pixel divided by stellar flux). While the flux cube would have the star’s spectral features imprinted on it, and thus need to be calculated at higher spectral resolution, the contrast cube does not, allowing for lower spectral resolution, quicker calculations, and substantially smaller file sizes.

We note that in a handful of anomalous cases, a small number of pixels in high-zodi disks can converge on the final flux very slowly as we increase $n_{\rm z}$ and $n_{\rm xy}$, oscillating about a final value. This behavior eventually dominates available memory and overall runtime. To mitigate this issue, we limit $n_{\rm z} \le 32768$ and $n_{\rm xy} \le 32$, regardless of the desired tolerance. To track this issue, exoVista outputs a map of the approximate precision achieved in each pixel, set equal to the absolute value of the fractional difference from the previous calculation. Users can reference this map to approximate the numerical noise in the debris disk image at all wavelengths.

Next, we evolve the orbits of the star and planets, recording the barycentric and heliocentric positions and velocities of all bodies as a function of time over the interval $t_{\rm max}$, resolved into timesteps of $\Delta t$. By default, $t_{\rm max}=5$ years and $\Delta t=10$ days. We use a Bulirsch-Stoer n-body integrator to calculate positions and velocities of all bodies to a tolerance of $10^{-12}$ \citep{starkkuchner2008}. We include only gravitational forces and neglect the gravitational influence of the debris disk on the rest of the system. Because all disks are assumed to be azimuthally symmetric, we do not include the dust particles in the integration, as the disk's appearance is assumed to remain fixed.

We then calculate each planet's reflected light contrast as a function of wavelength for each time step/position. To do so, we load each planet's geometric albedo spectrum at its native resolution, typically much higher than the default $R=300$, calculate the stellar spectrum at the same wavelengths, and determine the planet's reflected light flux at the higher native resolution. We then bin down to the desired output resolution, taking care to calculate the planet's flux at the exact transition wavelength between the lower resolution spectral bins. To convert to contrast, we divide by the stellar flux at the lower output resolution, such that users can retrieve the planet spectrum calculated at high fidelity simply by multiplying by the output stellar flux.

Finally, we produce an output scene file in .fits format. This file includes all of the stellar parameters generated by {\tt load\_stars}, the stellar position, velocity, and flux spectrum as a function of time, all of the planet parameters generated by {\tt generate\_planets}, the position, velocity, and contrast spectrum of all planets as a function of time, and the disk contrast data cube.

The {\tt generate\_scene} module is the most numerically taxing step of the exoVista code. Whereas previous modules execute in seconds, this step requires $\sim$1 day of execution on a 32 core machine to generate $\sim$8000 astrophysical scenes. The vast majority of the time is spent generating the debris disk image. Typical run times are $\sim$4 minutes per planetary system using a single core.

\subsection{{\tt load\_scene}}

The outputs of exoVista are .fits files that adhere to a specific format. The format is described in detail in the public documentation. To simplify the use of the data sets we have included a sample module called {\tt load\_scene} in both IDL and Python.  {\tt load\_scene} provides an example of how to convert the data products into an astrophysical data cube at a given time. {\tt load\_scene} opens the desired .fits file, interpolates all positions, velocities, and spectra to the desired epoch, and returns these quantities along with a disk flux data cube, allowing the user to easily generate an image by convolving with a user-supplied PSF.

\section{Results} \label{sec:results}

\subsection{Example analyses}

\begin{figure}[h]
	\centering	
	\includegraphics[width=7in]{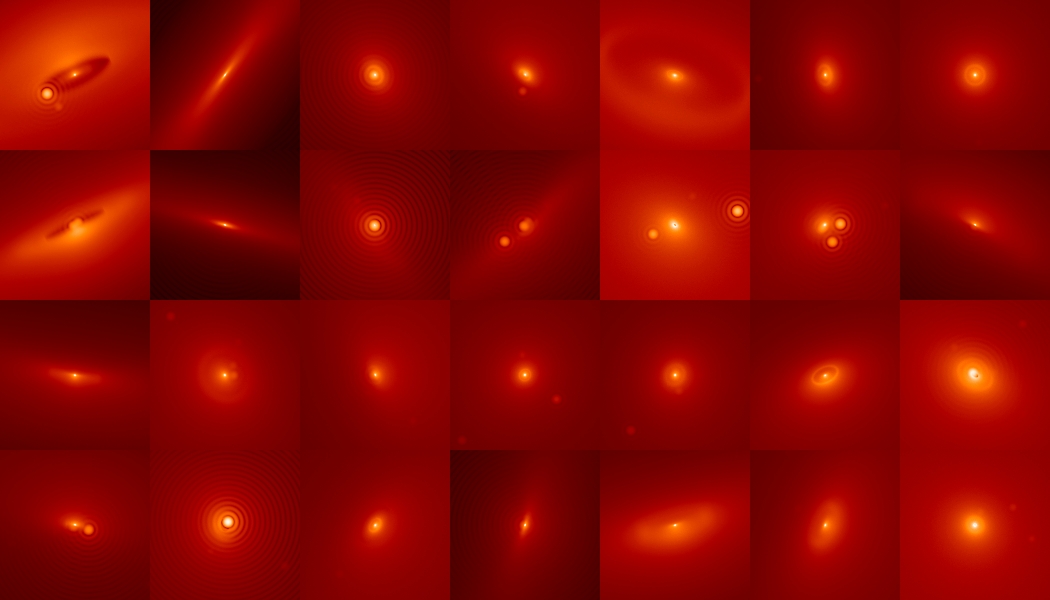}
	\caption{Example V band images of 28 random planetary systems generated around stars within 50 pc by exoVista. For illustrative purposes, the stellar flux is removed and the disk is not convolved with the PSF. \label{fig:images}}
\end{figure}

Figure \ref{fig:images} illustrates 28 random V band scenes generated by exoVista via the {\tt load\_scene} module. Each image is individually normalized, measures 0.3$\arcsec$ across, and neglects the stellar contribution. We modeled point sources using a 10 m telescope with an Airy pattern PSF. For illustrative purposes, to highlight the fidelity of the disk model, we did not convolve the disk with the PSF. A movie showing the systems in motion over a 5 year time span can be download at {\tt https://www.starkspace.com/permanent/exovista.mov}. Some disks show central clearings associated with massive planets due to the truncation feature of our disk model. Others show ring-like features associated with the parent body belt. Some systems exhibit bright inner planets, as evidenced by bright Airy rings extending to large distances. In highly inclined systems, these bright inner planets can create a stroboscopic effect as they quickly orbit from crescent to gibbous phase. The scenes shown in Figure \ref{fig:images} can be used to study a wide range of aspects of a future direct imaging mission, from how well we can model and subtract exozodi, to planet-planet confusion and orbit determination, to survey optimization. 

\begin{figure}[h]
	\centering	
	\includegraphics[width=4in]{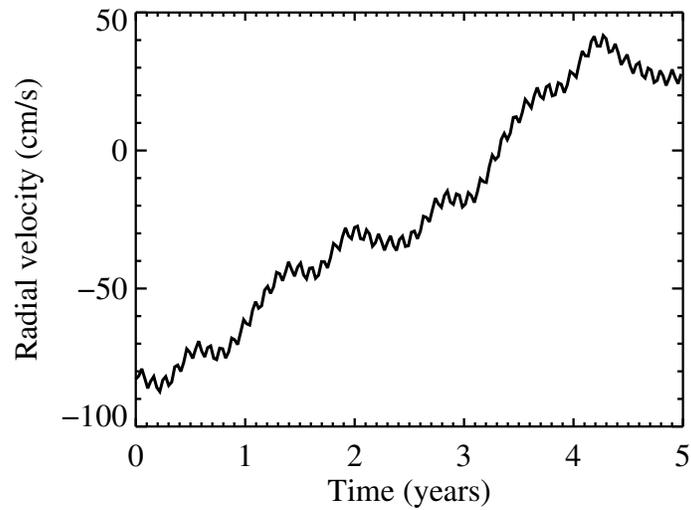}
	\caption{Example stellar radial velocity information extracted from a planetary system model generated by exoVista. Stellar noise/activity are not included.\label{fig:rv_plot}}
\end{figure}

\begin{figure}[h]
	\centering	
	\includegraphics[width=4in]{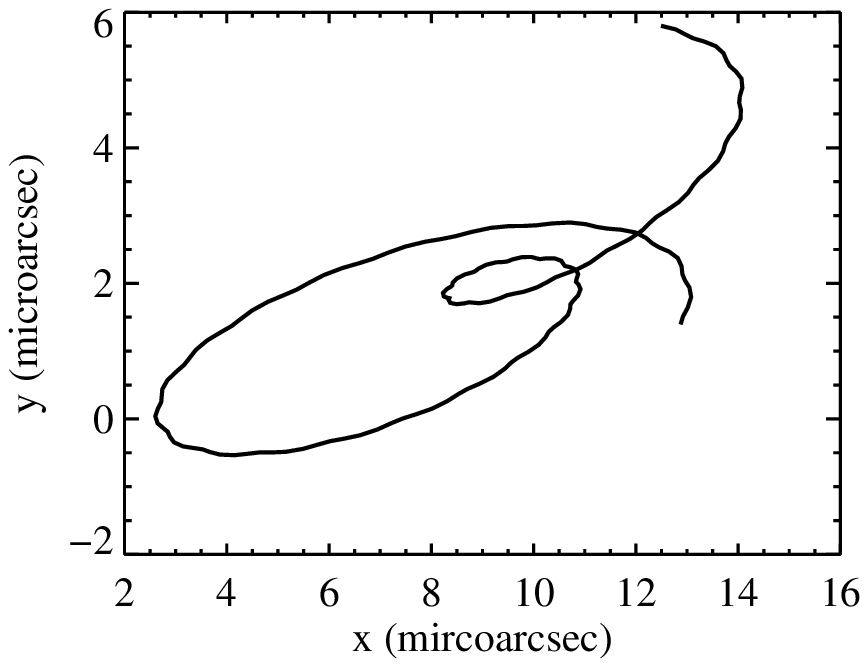}
	\caption{Example astrometric position extracted from a planetary system model generated by exoVista. The gravitational influence of companions stars and photometric asymmetries within the system are not included.\label{fig:astrometric_plot}}
\end{figure}

We can also use the outputs of exoVista to investigate other exoplanet detection methods. Figures \ref{fig:rv_plot} and \ref{fig:astrometric_plot} plot the barycentric radial velocity and astrometric position of a single star from our models, which clearly exhibit the signatures of several planets within the system. We note that exoVista does not currently include stellar noise, the gravitational influence of companion stars, or photometric asymmetries within the planetary system. As a result, these simulated signals are relatively crude and any detailed study of an RV or astrometric survey applied to these models would need to include such effects. 

\begin{figure}[h] 
	\centering	
	\includegraphics[width=4in]{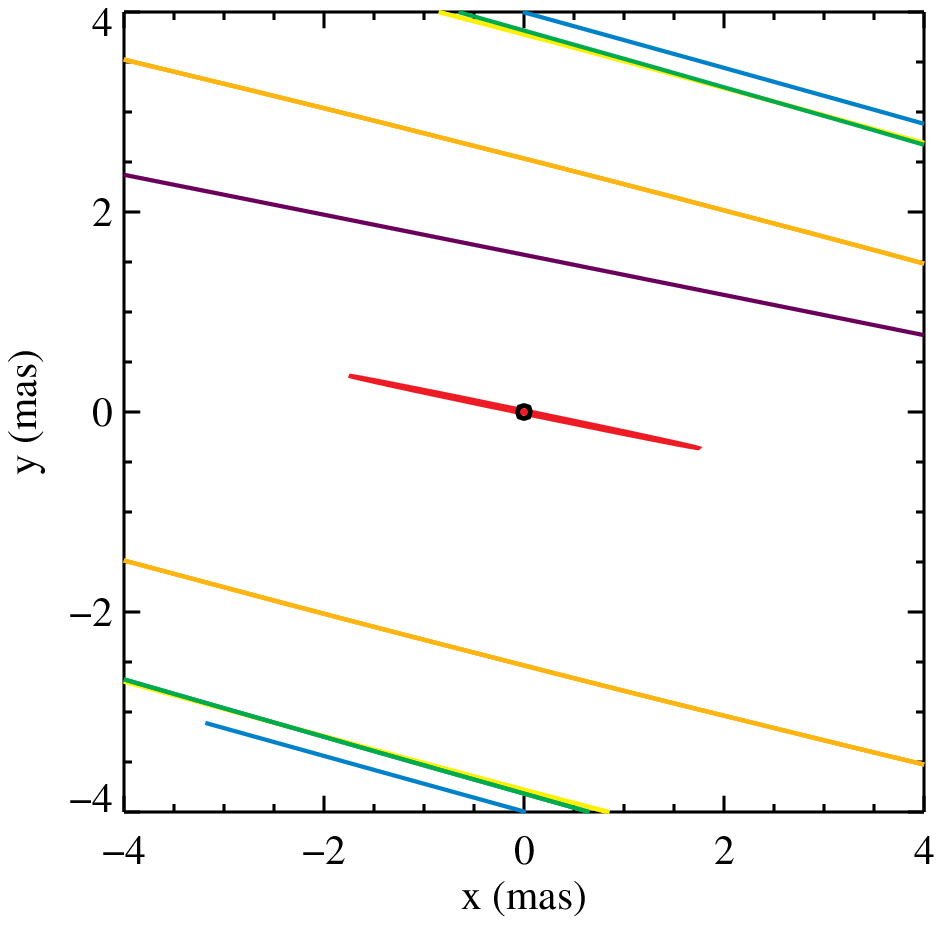}
	\caption{Illustration of the positions of several planets in a planetary system over a timespan of 5 years generated by exoVista. The inner planet (red) transits in front of the host star (black).\label{fig:transit_plot}}
\end{figure}

Figure \ref{fig:transit_plot} shows the position information plotted for multiple planets in a single near-edge-on system, each shown with a separate color. We plot the stellar radius at the center (black) to show that the innermost planet in this system (red) is predicted to transit. By comparing the planet radius with the stellar radius, users can generate transit light curves from the exoVista models. While the majority of transiting planets in the dataset should be reliably detected by interpolating the default time step of 10 days, grazing transits may require a finer time step.

\begin{figure}[h] 
	\centering	
	\includegraphics[width=4in]{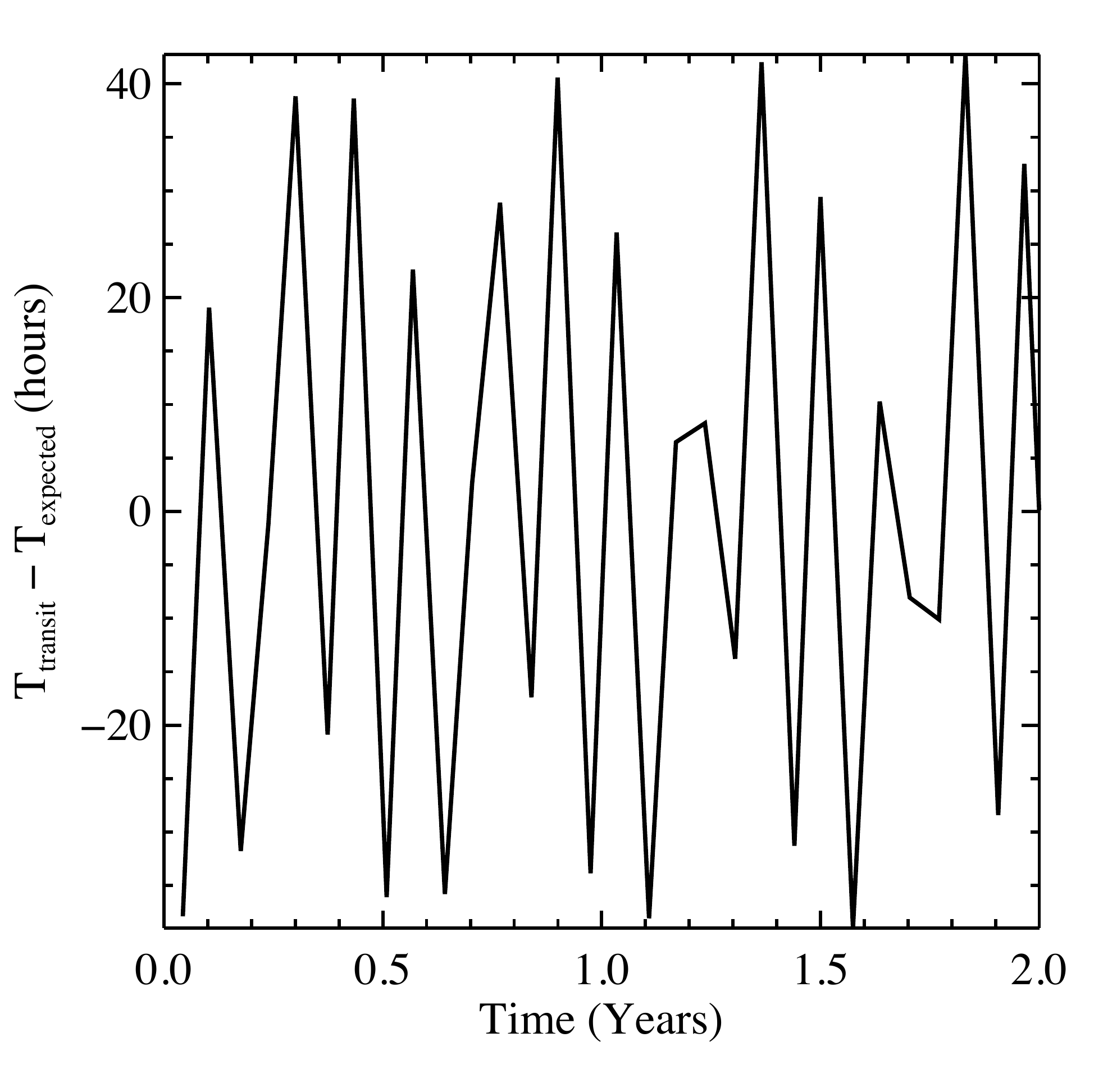}
	\caption{Example TTV plot for a transiting planet generated by exoVista. The gravitational influence of other planets impacts the time of transit ingress.\label{fig:ttv_plot}}
\end{figure}

ExoVista can also be used to perform rudimentary transit timing variation (TTV) analyses. Figure \ref{fig:ttv_plot} shows variations in the transit time of a modeled planet that transits, plotted as the difference of the ingress time from the average. Here we have simply interpolated the planet's position to estimate the moment of ingress. Detailed TTV analyses may require a finer time step in the dataset.

\subsection{Access}

The exoVista source code is publicly available for download at \url{https://www.starkspace.com}. The exoVista code is written in a combination of IDL and C. The provided documentation describes the installation process for several use case scenarios.

We expect that most users of exoVista may be more interested in the output fits-formatted models than the source code. Therefore, we have pre-generated planetary systems for 2258 stars within $30$ pc from the LUVOIR/HabEx master target list, as a preliminary exoVista ``universe." Because each realization of a planetary system is random, and a single instance does not capture the stochastic nature of these calculations, we repeated this process a total of 12 times. All 12 sets of 2258 planetary systems, which we refer to as the DEC21 data set, are available for download via the Goddard EMAC web server. Users can access the files via a browser-based file viewer at \url{https://ckan-files.emac.gsfc.nasa.gov/minio/exovista/}. At this same location we provide a {\tt README.txt} file that gives instructions on how to search for planetary systems that meet some parametric criteria and then retrieve those matching data sets. The data files can be read using the IDL or Python versions of the {\tt load\_scene} module in the exoVista source code distribution.

\section{Summary} \label{sec:summary}

ExoVista provides a new suite of modeled planetary systems that can be used to simulate a broad range of exoplanet detection methods. ExoVista randomly assigns planets to an input catalog of stars based on occurrence rates inferred from the \emph{Kepler} mission and RV survey observations and checks for dynamic stability, generates debris disks quasi-self-consistently with the underlying planetary system and measured exozodi distributions, and evolves the system using an n-body integrator. The resulting outputs include the position, velocity, and flux of all bodies as well as a contrast data cube of the debris disk, allowing for simulated study with the direct imaging, astrometric, transit, and RV methods.

\acknowledgments

We thank Neil Zimmerman and Jens Kammerer for translating the {\tt load\_scene} IDL code to Python, Aki Roberge for providing the geometric albedo files from the Haystacks exoplanet models, Giada Arney for providing geometric albedo files for Earth-like exoplanets, and Bertrand Mennesson for generating and providing the LBTI HOSTS best fit free-form exozodi distribution. This research made use of the NASA Exoplanet Modeling and Analysis Center (EMAC), which is funded by the NASA Planetary Science Division’s Internal Scientist Funding Model.






\bibliography{bibliography.bbl}
\bibliographystyle{aasjournal}

\begin{deluxetable}{p{4.5cm}p{0.8cm}p{1.5cm}p{11cm}}
\tablecolumns{4}
\tabletypesize{\scriptsize}
\tablecaption{ Parameters/Inputs Used by exoVista \label{table:parameters}}
\tablehead{\colhead{Parameter} & \colhead{Symbol} & \colhead{Default} & \colhead{Description/Source/Notes}\vspace{-0.3cm}\\
 $ $ & $ $ & \colhead{Value}\vspace{-0.5cm} & $ $\\}
\startdata
\hline
\hline
$ $ & $ $ & $ $ & Stellar Parameters \\
\hline
Input catalog & & & HabEx \& LUVOIR master target lists \citep{stark2019}, only MS stars within 50 pc, incomplete for $m_V\gtrsim8$\\
Distance & $d$ & $ $ & \emph{Gaia} TGAS \& DR2 \citep{gaiadr1,marrese2018}\\
Diameter & $ $ & $ $ & Analytic fits to B-V from \citet{boyajian2014}\\
Mass & $M_{\star}$ & $ $ & Interpolated from B-V via \citet{pecaut2013}\\
Effective Temperature & $ $ & $ $ & Interpolated from B-V via \citet{pecaut2013}\\
Luminosity & $L_{\star}$ & $ $ & Equation 9 of \citet{torres2010} with color correction from \citet{pecaut2013}\\
Photometry & $ $ & $ $ & \citet{perryman1997} and SIMBAD\\
Spectra & $ $ & $ $ & \citet{castelli2004} scaled to V band flux; model selected by stellar effective temperature, luminosity, and log(g)\\
Binarity & $ $ & $ $ & Separation and differential magnitude from Washington Double Star catalog\\
\hline
\hline
$ $ & $ $ & $ $ & Planet Parameters \\
\hline
System inclination & $i$ & $[0,\pi)$ & Inclination of the system midplane, drawn uniformly in $\cos{i}$ \\
System position angle & PA & $[0,2\pi)$ & Position angle of the system midplane, drawn uniformly \\
Occurrence rates & $ $ & $ $ & Planet occurrence rates in $(M,a_{L_{\star}})$ space \citep{dulz2020} \\
Semi-major axis & $a$ & $ $ & Drawn randomly from occurrence rates and checked for stability via $\Delta>6$ \\
Eccentricity & $e$ & $0$ & Non-zero range can be specified \\
Inclination & $ $ & $[0,5^{\circ}]$ & Drawn uniformly, relative to system midplane \\
Long. of ascending node & $ $ & $[0,2\pi)$ & Drawn uniformly \\
Argument of pericenter & $ $ & $[0,2\pi)$ & Drawn uniformly \\
Mean anomaly & $ $ & $[0,2\pi)$ & Drawn uniformly \\
Mass & $M$ & $ $ & Drawn randomly from occurrence rates and checked for stability via $\Delta>6$ \\
Radius & $R$ & $ $ & Singular value of $R$ for each $M$ via \citet{chenkipping2017} \\
Albedo & $ $ & $ $ & Wavelength dependent, drawn from planet models of \citet{roberge2017} \\
Phase function & $ $ & Lambertian & Wavelength-independent phase function applied to all planets \\
\hline
\hline
$ $ & $ $ & $ $ & Disk Parameters \\
\hline
Number of disk components & $ $ & 2 & Warm \& cold components by default, third component possible\\
Density (warm component) & $z_{\rm warm}$ & $ $ & Zodi level drawn from LBTI best fit free-form distribution \citep{ertel2020}\\
Density (cold component) & $z_{\rm cold}$ & $[0.2,5] \, z_{\rm warm}$ & Drawn uniformly in log($z_{\rm cold}/z_{\rm warm}$)\\
Disk scale height & $H/r_0$ & $[0.03,0.2]$ & Drawn uniformly for each component, modeled as a Gaussian dist. in $z$\\
Gaussian ring distance (warm component) & $r_0$ & $[0.5,5]$ AU & Circumstellar distance of Gaussian ring, drawn uniformly in log($r$) from stable regions\\
Gaussian ring distance (cold component) & $r_0$ & $[5,50]$ AU & Circumstellar distance of Gaussian ring, drawn uniformly in log($r$) from stable regions\\
Gaussian ring width & $\Delta r / r_0$ & $[0.05,0.3]$ & Smaller of the range maximum and the width allowed by stability is chosen\\
Gaussian ring size dist. & $ $ & Dohnanyi & Dust grain size distribution within the Gaussian ring\\
Halo distance & $ $ & $r_0 + \Delta r$ & Circumstellar distance of the inner edge of the halo\\
Halo density power law & $ $ & -1.5 & Radial exponent of the disk density in the halo\\
Halo size distribution & $ $ & Dohnanyi & Dust grain size distribution in the halo\\
Collision time & $t_{\rm coll}$ & $ $ & Analytically estimated from orbit time and number of zodis\\
Truncation mass & $ $ & $100$ M$_{\Earth}$ & Minimum mass of planet required to invoke ejected dust model\\
Truncation radius & $r_{\rm truncate}$ & $1.1r_{\rm ap}$ & Outer edge of the ejected dust model, where $r_{\rm apo}$ is the apastron distance of the outermost planet interior to $r_0$ with $M > 100$ M$_{\Earth}$\\
Truncation power law & $ $ & 3 & Exponent of the radial power law for the ejected dust model\\
Blowout size & $ $ & $0.5$ $\mu$m & Minimum grain size of all components, independent of spectral type\\
SPF & $ $ & $ $ & Linear combination of three HG functions\\
HG Weights & $w_1$ & $[0.6,0.9]$ & Weight assigned to first HG function, drawn uniformly\\
$ $ & $w_2$ & $[0.1,1-w_1]$ & Weight assigned to second HG function, drawn uniformly\\
$ $ & $w_3$ & $1-w_1-w_2$ & Weight assigned to third HG function, drawn uniformly\\
HG asymmetries & $g_1$ & [0.85,0.98] & HG asymmetry parameter for first HG function, drawn uniformly\\
$ $ & $g_2$ & [0.5,0.7] &  HG asymmetry parameter for second HG function, drawn uniformly\\
$ $ & $g_3$ & [-0.2,0.2] &  HG asymmetry parameter for third HG function, drawn uniformly\\
\hline
\hline
\enddata
\vspace{-0.8cm}
\tablecomments{blah blah blah}
\end{deluxetable}
\end{document}